\documentclass[graybox]{svmult}
\usepackage{mathptmx}       
\usepackage{helvet}         
\usepackage{courier}        
\usepackage{makeidx}         
\usepackage{graphicx}        

\usepackage{multicol}        
\usepackage[bottom]{footmisc}

\usepackage{cite}

\usepackage{fancyvrb}       

\usepackage{amsmath}
\usepackage{amssymb}
\usepackage{latexsym}
\usepackage{color,graphicx}
\usepackage{subfig}

\usepackage[color=yellow]{todonotes}

\newcommand{\CurlSymb}{\text{curl}}

\newcommand{\Curl}[2][]  {\text{\CurlSymb}{#1}\,{#2}}

\usepackage[linesnumbered,ruled]{algorithm2e}
\usepackage{algcompatible}

\begin{document}

\title*{POD-based reduced-order model of an eddy-current levitation problem}

\author{MD Rokibul Hasan
  \and Laurent Montier
  \and Thomas Henneron
  \and Ruth V. Sabariego}
\institute{
  MD Rokibul Hasan \and Ruth V. Sabariego \at
  KU Leuven, Dept. Electrical Engineering (ESAT), Leuven \& EnergyVille, Genk, Belgium, \
  \email{rokib.hasan@esat.kuleuven.be, ruth.sabariego@esat.kuleuven.be}
  \and
  Laurent Montier \and Thomas Henneron \at
  Laboratoire d'Electrotechnique et d'Electronique de Puissance, Arts et Metiers ParisTech, Lille, France,
  \email{laurent.montier@ensam.eu, thomas.henneron@univ-lille1.fr}
}

\maketitle

\abstract*{The accurate and efficient treatment of eddy-current problems with movement is still a challenge. Very few works applying reduced-order models are available in the literature. In this paper, we propose a proper-orthogonal-decomposition reduced-order
  model to handle these kind of motional problems. A classical magnetodynamic finite element formulation
based on the magnetic vector potential is used as reference and to build up the reduced models. Two approaches are proposed.   
The TEAM workshop problem 28 is chosen as a test case for validation.
Results are compared in terms of accuracy and computational cost.
} 

\abstract{The accurate and efficient treatment of eddy-current problems with
  movement is still a challenge. Very few works applying reduced-order
  models are available in the literature. In this paper, we propose a
  proper-orthogonal-decomposition reduced-order model to handle these kind
  of motional problems. A classical magnetodynamic finite element
  formulation based on the magnetic vector potential is used as reference and to build up
the reduced models. Two approaches are proposed. The TEAM workshop problem 28 is chosen as a test case for validation.
Results are compared in terms of accuracy and computational cost.
}

\section{Introduction}

The finite element (FE) method is widely used and versatile for accurately modelling
electromagnetic devices accounting for eddy current effects, non-linearities, movement,...
However, the FE discretization may result in a
large number of unknowns, which maybe extremely expensive in terms of computational time and memory.
Furthermore, the modelling of a movement requires either remeshing or ad-hoc techniques. Without being exhaustive, it is worth mentioning: the hybrid
finite-element boundary-element (FE-BE) approaches~\cite{Sabariego2004}, the
sliding mesh techniques (rotating machines)~\cite{Boualem1998} or the mortar FE approaches~\cite{Rapetti2010}.

Physically-based reduced models are the most popular approaches for
efficiently handling these issues. They extract physical parameters (inductances, flux
linkages,...) either from simulations or measurements and construct
look-up tables covering the operating range of the device at hand~\cite{2005lee_RO_team28,Liu2007}. 
Future simulations are performed by simple interpolation, drastically
reducing thus the computational cost. However, these methods depend highly on the
expert's knowledge to choose and extract the most suitable parameters. 

Mathematically-based reduced-order (RO) techniques are a feasible
alternative, which are gaining interest in electromagnetism~\cite{Schilders2008}.
RO modelling of static coupled system has already been implemented
in~\cite{Yue2015,Banagaaya2015}.
Few RO works have addressed problems with movement (actuators, electrical
machines, etc.)~\cite{Albunni2008,Henneron2014,Sato2015}.

In~\cite{Albunni2008}, authors consider a POD-based FE-BE model electromagnetic device
comprising nonlinear materials and movement. Meshing issues are avoided but
the system matrix is not sparse any more, increasing considerably the cost of
generating the RO model. In~\cite{Henneron2014}, a magnetostatic POD-RO model of a
permanent magnet synchronous machine is studied. A locked step approach is used, so the mesh and associated number of unknowns remains constant. 
A POD-based block-RO model is proposed in~\cite{Sato2015,Schmidthausler2014}, where the domain is split in linear
and nonlinear regions and the ROM is applied only to the linear part.

\begin{figure}
\centering
\parbox{5cm}{
\includegraphics[width=5cm]{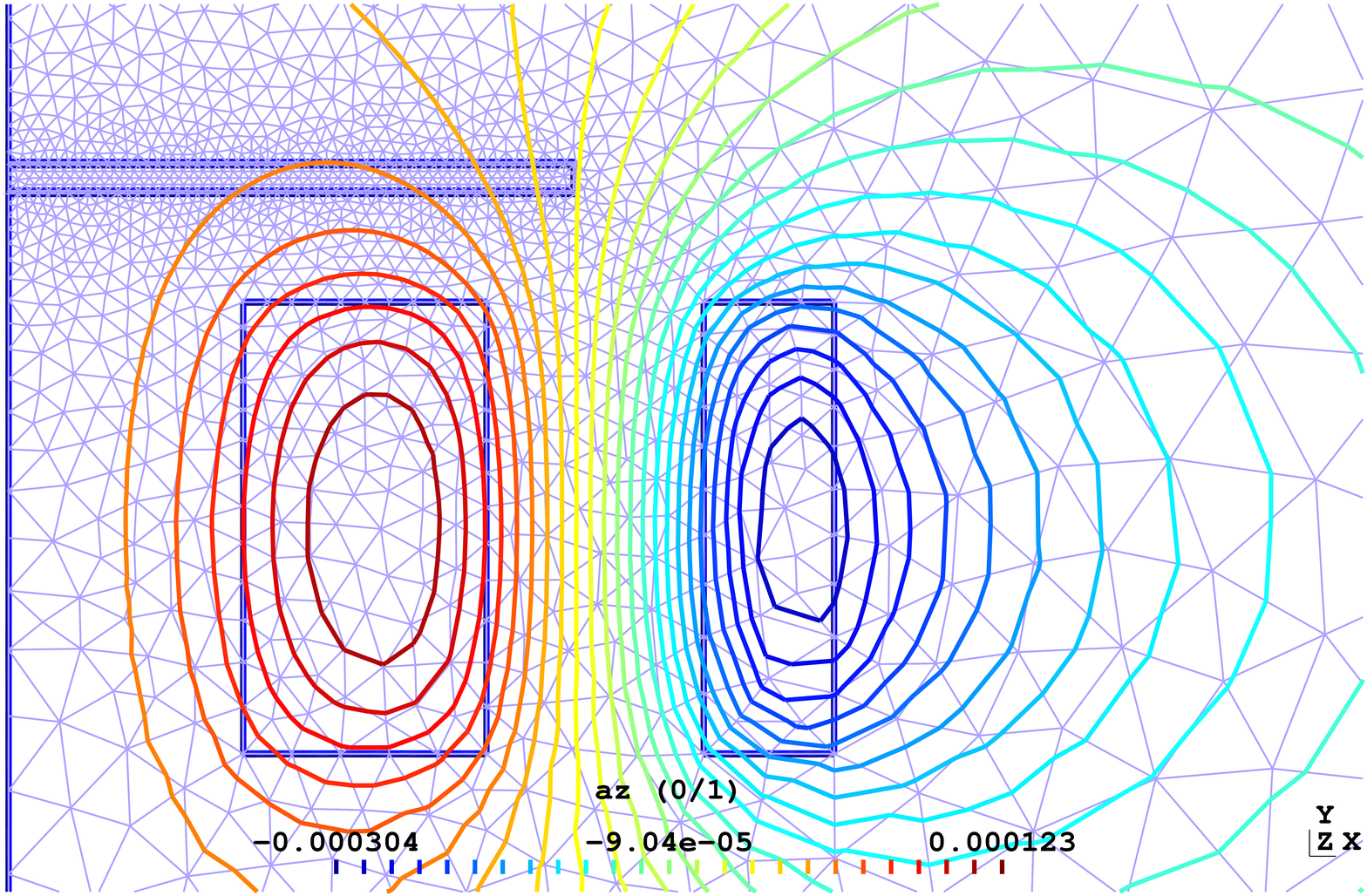}
\label{fig:t28_1stmodel}}
\qquad
\begin{minipage}{5cm}
\includegraphics[width=5cm]{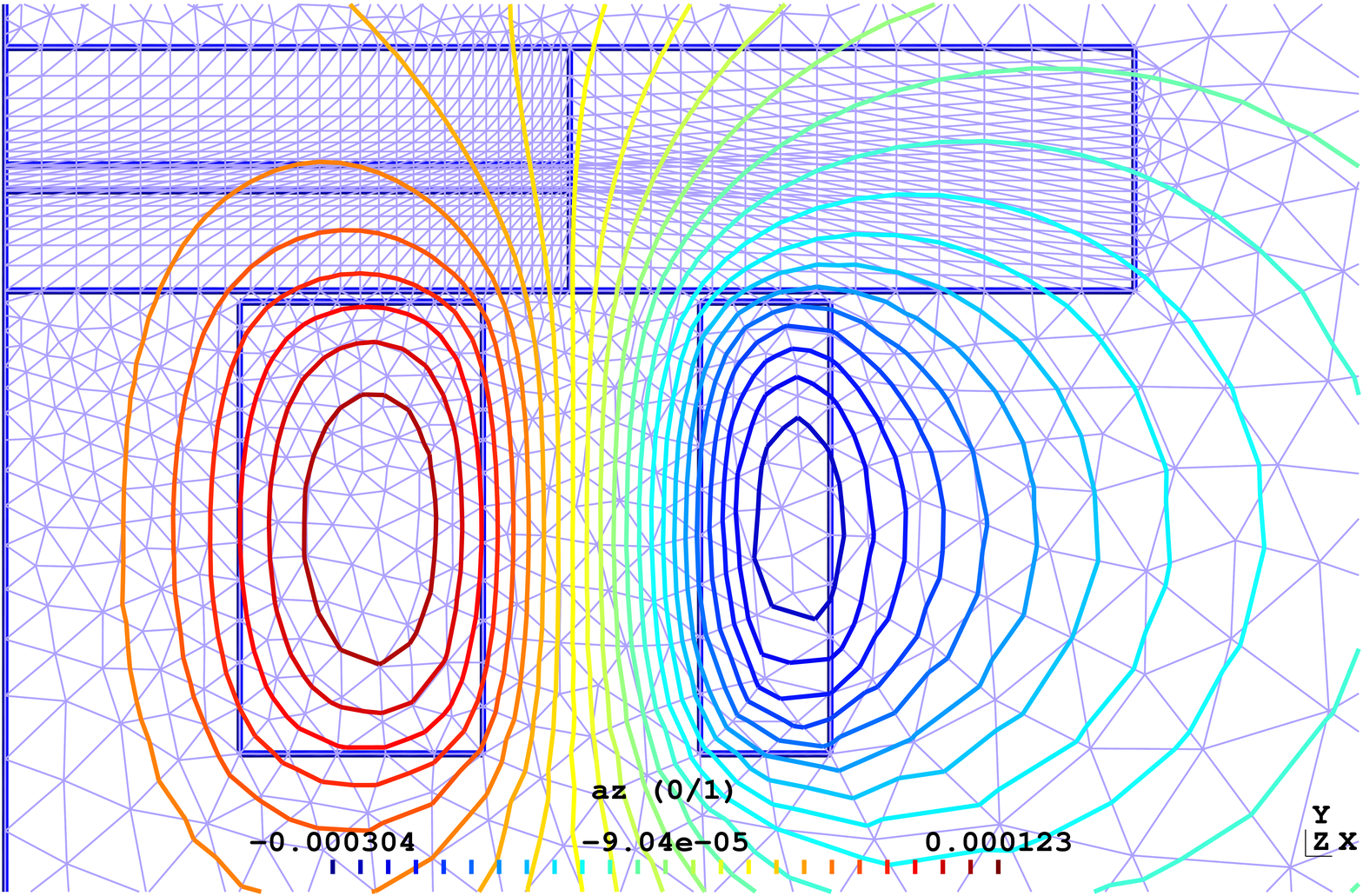}
\label{fig:t28_2ndmodel}
\end{minipage}
\caption{2D axisymmetric mesh of TWP28: aluminium plate above two concentric
  coils (12.8\,mm clearance). Real part of the magnetic flux density. Left: automatic remeshing of the full domain, Right: mesh deformation of sub-domain around plate with nodes fixed at it's boundaries (except axes).}
\label{fig:t28_both_model}
\end{figure}

In this paper, we consider a POD-based FE model of a levitation problem, namely the Team
Workshop problem 28 (TWP28)~\cite{TEAM28, 2005lee_RO_team28} (a conducting plate
above two concentric coils, see Fig.~\ref{fig:t28_both_model}). 
The movement is modelled with two RO models based on: 1) FE with automatic remeshing of the
complete domain; 2) FE with constraint remeshing, i.e., localized deformation of the mesh around the
moving plate, hereafter referred to as mesh deformation. 
Both models are validated in the time domain and compared in terms of computational efficiency.

\section{Magnetodynamic levitation model}
Let us consider a bounded domain $\Omega=\Omega_c\cup\Omega_c^C \in \mathbb{R}^3$ with
boundary $\Gamma$. The conducting and non-conducting parts of $\Omega$ are
denoted by $\Omega_c$ and $\Omega_c^C$, respectively.
The (modified) magnetic-vector-potential ($a-$) magnetodynamic formulation (weak form of Amp\`ere's law) reads: 
find $a$, such that
\begin{align}
(\nu \, \Curl{a}, \Curl{a'})_{\Omega} + 
(\sigma \partial_t a, a')_{\Omega_c} +
\langle \hat{n} \times h, a' \rangle_{\Gamma}
=(j_s, a')_{\Omega_s}\,, \ \forall a'
\label{eqn:weak_TD}
\end{align}
with $a'$ test functions in a suitable function space;
$b (t)=\Curl{a (t)}$, the magnetic flux density; $j_s (t)$ a prescribed current
density and $\hat{n}$ the outward unit normal vector on $\Gamma$. 
Volume integrals in $\Omega$ and surface integrals on $\Gamma$ of the scalar
product of their arguments
are denoted by $(\cdot,\cdot)_\Omega$ and
$\langle\cdot,\cdot\rangle_\Omega$.
The derivative with respect to time is denoted by $\partial_t$.
We further assume linear isotropic and time independent materials with
magnetic constitutive law, so that the magnetic field is $h (t)=\nu b (t)$
(reluctivity $\nu$) and electric constitutive law, given by induced eddy current density $j (t)=\sigma e (t)$,
(conductivity $\sigma$) where, electric field $e (t)=- \partial_t a
(t)$. Assuming a rigid $\Omega_c$ (no deformation) and a purely
translational movement (no rotation, no tilting), the electromagnetic force
appearing due to the eddy currents in $\Omega_c$ can be modelled as a global
quantity with only one component (vertical to the plate). If $\Omega_c$ is
non-magnetic, Lorentz force can be used:

\begin{equation}
F_{em}(t)= \int_{\Omega_c}  j(t) \times b(t)\; \mathrm{d}{\Omega_c} = \int_{\Omega_c} -\sigma \partial_t a (t) \times \Curl{a (t)}\; \mathrm{d}{\Omega_c}\,.
  \label{eqn:lorentz}
\end{equation}

The 1D mechanical equation governing the above described levitation problem reads:
\begin{equation}
\begin{aligned}
   m\,\partial_t v(t)+ \xi v(t) + k y(t) + mg = F_{em}(t)\,
  \label{eqn:mechanical}
  \end{aligned}
\end{equation}
where unknown $y(t)$ is the center position of the moving body in the vertical direction,
$v(t)= \partial_t y(t)$ is the velocity of the moving body, 
$m$ is the mass of the moving body, $g$ is the acceleration of gravity,
$\xi$ is the scalar viscous friction coefficient,
$k$ is the elastic constant. 
We apply the backward Euler method to solve~\eqref{eqn:mechanical}. The moving body displacement of system~\eqref{eqn:mechanical} results from the ensuing electromagnetic force generated by system~\eqref{eqn:weak_TD} and thus affects the geometry. Given that, the dynamics of the mechanical equation is much slower than the electromagnetic equation, if the time-step is taken sufficiently small, one can decouple the equations. Under this condition, the electromagnetic and mechanical equations can be solved alternatively rather than simultaneously by the weak electromechanical coupling algorithm of~\cite{Henrotte1994}. We adopt this approach.
\section{POD-based model order reduction}

The proper orthogonal decomposition (POD) is applied to reduce the matrix system
resulting from the FE discretisation of~\eqref{eqn:weak_TD}:
\begin{equation}
A\partial_tx(t)+B x(t)=C(t)\,.
\label{eqn:system}
\end{equation}
where $x(t)\in \mathbb{R}^{N \times 1} $ is the time-dependent column vector of $N$ unknowns, $A$, $B$
$\in \mathbb{R}^{N \times N}$ are the matrices of coefficients and $C(t)\in \mathbb{R}^{N \times 1}$ is the source column
vector.
Furthermore, the system~\eqref{eqn:system} is discretized in time by means
of the backward Euler scheme. A system of algebraic equation is obtained for
each time step from $t_{k-1}$ to $t_k= t_{k-1} +\Delta t $, $\Delta t$ the step size.
The discretized system reads:
\begin{equation}
\left[A_{\Delta t} + B \right] x_k= A_{\Delta t} x_{k-1} + C_k\,
\label{eq:time_euler}
\end{equation}
with $A_{\Delta t}=\frac{A}{\Delta t}$, $x_k=x(t_k)$ the solution at instant $t_k$, $x_{k-1}=x(t_{k-1})$ the solution at instant
$t_{k-1}$, $C_k$ the right-hand side at instant $t_k$.

In RO techniques, the solution vector $x (t)$ is approximated by a
vector $x^r (t)$ $\in \mathbb{R}^{M\times 1}$ within a reduced subspace
spanned by $\Psi\in \mathbb{R}^{N \times M}$, $M \ll N$,
\begin{equation}
x(t) \approx \Psi x^r (t)\, , 
\label{eq:reduced_freq}
\end{equation}
with $\Psi$ an orthonormal projection operator 
generated from the time-domain full solution $x (t)$
via snapshot techniques~\cite{Sato2013}.

Let us consider the snapshot matrix, $S=[x_1, x_2, \ldots, x_M ]\in \mathbb{R}^{N \times M}$ from the set of
solution $x_k$ for the selected number of time steps.
Applying the singular value decomposition (SVD) to $S$ as,
\begin{equation}
S=\mathcal{U} \Sigma \mathcal{V}^T\,.
\label{SVD}
\end{equation}
where $\Sigma$ contains the singular values, ordered as $\sigma_1 > \sigma_2>\ldots>0 $. We consider $\Psi=\mathcal{U}^r\in \mathbb{R}^{N \times r}$, that  corresponds to the truncation ($r$ first
columns, which has larger singular values than a pre-defined error tolerance $\varepsilon$) with orthogonal matrices $\mathcal{U}\in \mathbb{R}^{N \times r}$ and $\mathcal{V}\in \mathbb{R}^{M \times r}$. 
Therefore, the RO system of~\eqref{eq:time_euler} reads
\begin{equation}
\left[ A^r_{\Delta t}+B^r\right] x^r_k= A^r_{\Delta t} x^r_{k-1} + C^r_k\,,
\label{eq:RO_time}
\end {equation}
with $A^r_{\Delta t}=\Psi^T A_{\Delta t} \Psi$, $B^r=\Psi^T B \Psi$ and
$C^r=\Psi^T C$ ~\cite{Hasan.Energycon2016}.

%

\subsection{Application to an electro-mechanical problem with movement}
\subsubsection{RO modelling with automatic remeshing technique}
In case of automatic remeshing, we transfer results from the source mesh$_{k-1}$ to the new target mesh$_{k}$ by means of a Galerkin projection, which is optimal in the $L_2$-norm sense~\cite{Dular2008}. Note that, this projection is limited to the conducting domain, i.e. the plate, as it is only there that we need to compute the time derivative. The number of  unknowns per time step $t_k$
varies and the construction of the snapshot matrix $S$ is not
straightforward. As the solution at $t_k$ is supported on its own mesh, the snapshot vectors $x_k$ have a different size. They have to be
projected to a common basis using a simple linear interpolation
technique before being assembled in $S$ and getting the projection operator $\Psi$.
The procedure becomes thus extremely inefficient.
  
\subsubsection{RO modelling with mesh deformation technique}
\begin{figure}
\centering
\includegraphics[width=8cm]{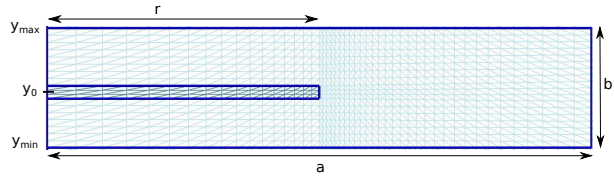}
\vspace{-0.25cm}
\caption{Sub-domain for deformation: plate position at $y_0 = 12.8$\,mm (initial mesh).}
\label{fig:box_plate}
\end{figure}

\begin{figure}
\centering
\includegraphics[width=8cm]{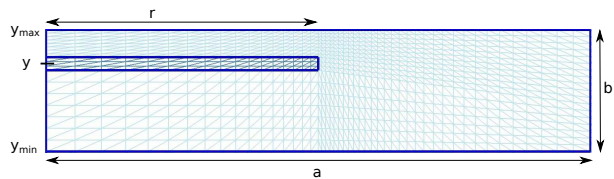}
\vspace{-0.25cm}
\caption{Sub-domain for deformation: plate position at $y = 20$\,mm. Mesh elements under the plate are expanded and above the plate are shrinked.}
\label{fig:box_plate_up}
\end{figure}

The automatic remeshing task is replaced by a mesh deformation technique, limited to a region around the moving body (see, e.g., the
box in Fig.~\ref{fig:box_plate}). Therefore, in this case, the remeshing is done by
deforming the initial mesh, which is generated with the conducting
plate placed at, e.g., $y_0$ (avoiding bad quality elements), see
Fig.~\ref{fig:box_plate}. The mesh elements only inside the sub-domain can be deformed (shrink/expand), see Fig.~\ref{fig:box_plate_up} and the nodes at the boundary of the sub-domain are fixed. The surrounding mesh does not vary. In our test case, we assume a vertical force (neglect the other two components) in~\eqref{eqn:lorentz}, therefore, the mesh elements only deform in the vertical direction and the nodes are fixed at the boundary of the sub-domain (not at the axes due to the axisymmetry). The size of the sub-domain (a$\times$b) is determined by the extreme positions of the moving body. In our validation example, the minimum position
(3.8\,mm) is given by the upper borders of the coils and the maximum
position (22.3\,mm) could be estimated by means of a circuital model,
e.g.~\cite{2005lee_RO_team28}. The number of unknowns
per time step remains now constant so the construction of matrix
$S$ is direct.
\vspace{-0.5cm}

\vspace{-0.5cm}

\vspace{-0.25cm}
\noindent
\begin{minipage}{\textwidth}
\noindent
\begin{minipage}[ht]{.50\textwidth}
  \begin{algorithm}[H]
    \caption{Automatic remeshing}
    \SetKwInOut{Input}{Input}
    \SetKwInOut{Output}{Output}
    \SetKwComment{Comment}{//}{}
    
    \Input{
      snapshot vectors $\{S_c\} \leftarrow \{x_k\}\in
      \mathbb{R}^{n(k)\times1}$
      \\
      time steps \\$\{t_k\}$, $k \in [1,...,K]$\\
      $A_{\Delta t}, \ B, \ C$,
      tolerance $\varepsilon$\\
      $m \leq n(k)$ snapshot vectors
    }
    \Output{displacement $y_k$}
    
    \BlankLine
    $y_0 = $ initial position,
    $\Delta y_0 = 0$
    \BlankLine
    \Comment{Time resolution}
    \For{$k \leftarrow 1$ \KwTo $K$}{
      \Comment{Magnetics}
      generate matrices
      $A_{\Delta t_k},\ B_k ,\ C_k $ \\
      find length of $C_k\in \mathbb{R}^{n(k)\times 1}$\\
      $S_p = \mathbf{0} \in \mathbb{R}^{n(k)\times m}$\\
      $S_p \leftarrow$ projection of $\{S_c\}$ to $n(k)$ rank subspace \\
      SVD of $S_p= \mathcal{U} \Sigma \mathcal{V}^T $\\    
      $\Psi_k = \mathcal{U}(:, 1\ldots r)$ with $r$ such that $ \sigma(i) / \sigma (1) > \varepsilon, \forall i \in [1\ldots r]$
        \\
        $
        \begin{aligned}
          A_{\Delta t_k}^{r}
          =& \Psi_k^T A_{\Delta t_k} \Psi_k,\\ B_k^{r}
          =& \Psi_k^T B_k \Psi_k,\\ C^r_k
          =& \Psi_k^T C_k
        \end{aligned}
        $
      \\
      solve $\left(A_{\Delta t_k}^{r} + B_k^{r}\right) x_{k}^r = C^r_k + A_{\Delta t_k}^{r} x_{k-1}^r$\\
      $x_k \approx \Psi_k x_{k}^r$\\
      compute force $F_k$\\
      \BlankLine
      \Comment{Mechanics}
      compute displacement $y_k$\\
      update $\Delta y_k = y_k-y_{k-1}$\\
      remesh with $y_k$ \\
    }
  \end{algorithm}
\end{minipage}
\hfill
\begin{minipage}[ht]{.50\textwidth}
  \begin{algorithm}[H]
    \caption{Mesh deformation}
    \SetKwInOut{Input}{Input}
    \SetKwInOut{Output}{Output}
    \SetKwComment{Comment}{//}{}
    
    \Input{
      snapshot matrix $S=[x_1, \ldots, x_m]\in \mathbb{R}^{n\times m}$,
      $x_k \in \mathbb{R}^{n\times1}$ \\
      time steps\\ $\{t_k\}$, $k \in [1,...,K]$\\
      $A_{\Delta t}, \ B, \ C $,
      tolerance $\varepsilon$\\
      $m \leq n$ snapshot vectors
    }
    \Output{displacement $y_k$}
    
    \BlankLine
    
    $y_0 = $ initial position, $\Delta y_0 = 0$\\
    get initial mesh
    \\
    SVD of $S = \mathcal{U} \Sigma \mathcal{V}^T $\\
    $\Psi_k = \mathcal{U}(:, 1\ldots r)$ with $r$ such that $ \sigma(i) / \sigma (1) > \varepsilon, \forall i \in [1\ldots r]$ \\
    \BlankLine
    \Comment{Time resolution}
    \For{$k \leftarrow 1$ \KwTo $K$}{
    \Comment{Magnetics}
    generate matrices $ A_{\Delta t_k},\  B_k , \ C_k $\\
    $
    \begin{aligned}
      A_{\Delta t_k}^{r}
      =& \Psi^T A_{\Delta t_k} \Psi,\\ B_k^{r}
      =& \Psi^T B_k \Psi,\\ C^r_k
      =& \Psi^T C_k
    \end{aligned}
    $
    \\
    solve
    $\left(A_{\Delta t_k}^{r} + B_k^{r}\right) x_{k}^r = C^r_k + A_{\Delta t_k}^{r} x_{k-1}^r$\\  
    $x_k \approx \Psi x_{k}^r$\\
    compute force $F_k$\\
    \BlankLine
    \Comment{Mechanics}
    compute displacement $y_k$\\
    update $\Delta y_k = y_k-y_{k-1}$\\
    deform mesh with $y_k$\\
  }
\end{algorithm}
\end{minipage}
\end{minipage}

\section{Application example}
We consider TWP28: an electrodynamic levitation device consisting of a conducting cylindrical aluminium 
plate ($\sigma = 3.47\cdot10^7 $ S/m, $m=0.107 $ Kg, $ \xi=1$) above two coaxial
exciting coils. The inner and outer coils have 960 and 576 turns
respectively. Note that, if we neglect the elastic force, the equilibrium is reached when the $F_{em}$ is $1$N.
At $t=0$, the plate rests above the coils at a distance of 3.8\,mm. 
For $t\geq0$, a time-varying sinusoidal current (20\,A, $f=50$\,Hz) is imposed, same
amplitude, opposite directions~\cite{TEAM28}. Assuming a translational movement (no rotation and tilting) we can use an axisymmetric  model.
A FE model is generated as reference and origin of the RO models.
We have time-stepped 50 periods (100 time steps per period and step size
0.2\,ms), discretization that ensures accuracy and avoids degenerated mesh
elements during deformation.

\subsection{RO modelling with automatic remeshing full domain}
\begin{figure}
  \begin{center}
    \includegraphics[height=5cm]{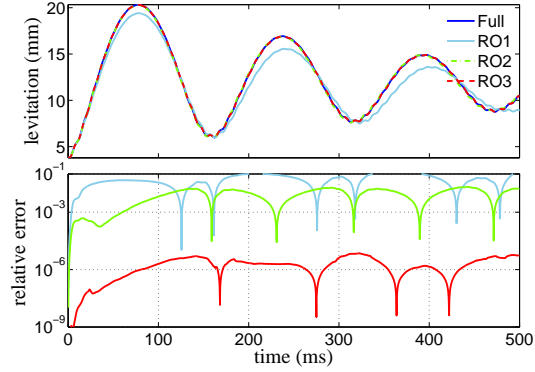}
  \end{center} 
  \caption{Displacement (up) and relative error (down) between full and RO models.}
  \label{fig:full_ROM} 
\end{figure}

In case of full domain remeshing, the first 1500 time steps (300\,ms) of the simulation, that correspond to the first two peaks (2P) in Fig.~\ref{fig:full_ROM}, are included in the snapshot matrix.

Three POD-based RO models are constructed based on the $r$ number of first singular value modes greater than a prescribed error tolerance $\varepsilon$, that is set manually observing the singular values decay curve of the snapshot matrix (see in Table~\ref{algo1_table}). 
The smaller the prescribed $\varepsilon$, the bigger the size of the RO model will be (size of RO3$>$RO2$>$RO1).
\begin{table}
\caption{$L_2$-relative errors of RO models on levitation height for 2P (automatic remeshing).}
\centering
\begin{tabular}{ | c | c | c | c |}
   \hline
   RO models & $M$   & $\varepsilon$ & rel. error    \\ 
  \hline
   RO1  		&   	$1085$	&   $10^{-6}$ & $1.25\cdot10^{-1}$					     	 \\ 
   RO2   	&	$1403$	&	$10^{-11}$	& $1.03\cdot10^{-2}$		\\ 
   RO3  	    &	$1411$	&   $10^{-15}$	& $2.45\cdot10^{-6}$			 
   \\
   \hline
 \end{tabular}
\label{algo1_table}
\end{table}

The displacement and relative error of the full and RO models are shown in
Fig.~\ref{fig:full_ROM}.  
Accurate results have been achieved with the truncated basis models: RO2
and RO3, with fix size per time step $M=1403$ and 1411. This approach is
completely inefficient, as the maximum number of unknowns we have in the
full model is 1552.

\subsection{RO modelling with mesh deformation of a sub-domain}
The choice of the sub-domain to deform the mesh is a non-trivial task: it should be
as small as possible while ensuring a high accuracy.
From our reference FE solution~\cite{TEAM28}, by observing the minimum and maximum levitation
height of the plate, we fixed the sub-domain size along the $y-$axis between $y_{min}=1.3$\,mm and
$y_{max}=29.3$\,mm, distances measured from the upper border of the coils.
The size along the $x-$axis has a minimum equal to the radius of the plate,
i.e.\ $r= 65$\,mm. This value is however not enough due to fringing effects. 
We have taken different size along the $x-$axis: $1.5 r, 2 r, 3 r$ (97.5, 130 and 195\,mm), measured from the axis
(Fig.~\ref{fig:box_plate}).
The meshed boxes yield
1921, 1836 and 1780 number of unknowns.

The relative errors in time shown in Table~\ref{error_table} decrease with the increasing sub-domain lengths/box
sizes considered. We have therefore chosen to further analyse the RO results
obtained with a box length along $x$ of 195\,mm ($3 r$). The discretization is kept
constant for all RO models computation.
\begin{table}
\caption{$L_2$-relative errors of RO models on levitation height for 1P (mesh deform).}
\centering
\begin{tabular}{ | c | c | c | }
   \hline
   sub-domain lengths (mm) & $M =7$   & $M =35$     \\ 
  \hline
   97.5  		&   	$8.24\cdot10^{-2}$	&	$6.14\cdot10^{-4}$				     	 \\ 
   130   	&	$5.71\cdot10^{-2}$	&	$1.90\cdot10^{-4}$			\\ 
   195  	&	$4.53\cdot10^{-3}$	&	$3.73\cdot10^{-5}$			 
   \\
   \hline
 \end{tabular}
\label{error_table}
\end{table}

The first 800 time steps (160\,ms) of the simulation, that correspond to the first peak (1P), are taken in the snapshot matrix in order to generate the projection basis $\Psi$. In the snapshot matrix, the most important time step solutions are included, which found as optimum selection for approximating the full solution. Then the basis is truncated as $\Psi=\mathcal{U}^r$ ($r$ first
columns) by means of prescribed error tolerance
($\varepsilon=10^{-5},10^{-8}$). The basis are truncated for 1P to get RO models of size $M= 7$ and $35$.

\begin{figure}
  \begin{center}
    \includegraphics[height=5.5cm]{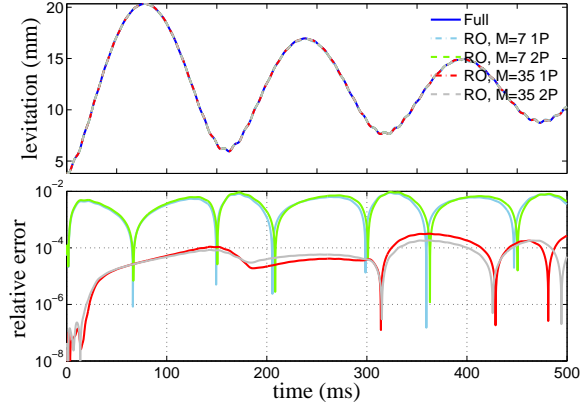}
  \end{center} 
\caption{Displacement (up) and relative error (down) between full and RO models for 195\,mm sub-domain length.}
\label{fig:disp_unk_1970} 
\end{figure}

From Fig.~\ref{fig:disp_unk_1970}, it can be observed that, RO model already
shows very good argument with only $M=7$ truncated basis, which is generated
from the snapshot matrix that incorporates first peak (1P). The accuracy of
RO models does not improve significantly with the addition of following
transient peaks (2P) into the snapshot matrix, but the accuracy certainly
improves with $M$. Hence, with $M=35$ the full and RO curves are
indistinguishable. The accuracy of the RO models can also be observed from
the $L_2$-relative errors figure.

With regard to the computation time (5000 time steps), the RO with $M=7$, can be solved less than an
hour, which is $3.5$ times faster than the full-domain automatic remeshing approach,
where the major time consuming part is to project the $\Psi$ on a same
dimensional basis as the system coefficient matrices, to reduce the system
in each time step.
Be aware that the computation is not optimized, performed on a laptop,
(Intel Core i7-4600U CPU at $2.10$\,GHz) without any parallelization.

\section{Conclusion}

In this paper, we have proposed two approaches for POD-based RO models to treat a
magnetodynamic levitation problem: automatic remeshing and mesh deformation of a sub-domain
around a moving body.
The RO model is completely inefficient with automatic remeshing technique, as the computational cost is nearly expensive as the classical approach.
The approach with sub-domain deformation to limit the influence of the
movement on the RO model construction has proved accurate and efficient (low
computational cost). 
We have shown results for three different sub-domain sizes, the bigger the
sub-domain the higher the accuracy. Further, computationally efficient RO
modelling of such parametric model is ongoing research.



\end{document}